\documentclass[%
nofootinbib,
 amsmath,amssymb,
 aps,
]{revtex4-2}

\usepackage{graphicx}
\usepackage[subrefformat=parens]{subcaption} 
\usepackage{dcolumn}
\usepackage{bm,colortbl,braket,slashed}%
\usepackage{hyperref}
  \hypersetup{colorlinks=true}

\newcommand{\1}{\mbox{1}\hspace{-0.25em}\mbox{l}} %

\begin{document}


\title{Vacuum decay and fermion total reflection by the Klein step}

\author{H. Nakazato}
	 \email{nakazato@hep.phys.waseda.ac.jp} 
\author{M. Ochiai}%
 \email{ochiai36@akane.waseda.jp}
\affiliation{%
 Department of Physics, Waseda University, 169-8555 Tokyo, Japan 
}%


\date{\today}

\begin{abstract} 
  The so-called Klein tunneling is re-examined within the framework of quantum field theory, but from a different point of view on the asymptotic states. We treat it as a one-dimensional scattering process of a fermion incident to a step potential and introduce asymptotic operators as appropriate $t = \pm \infty$ limits of the field operator responsible for the process. For the so-called Klein energy range, two asymptotic vacua naturally emerge which are defined as states annihilated by the asymptotic annihilation operators. They are related by a similarity transformation, which entails a vacuum decay and yields a vacuum decay constant. When a fermion with incident energy in the Klein region is injected to the step, it is shown to be reflected with probability one, accompanied by fermion--anti-fermion pairs that are vacuum decay products. 
\end{abstract}

\maketitle


\section{\label{Sec1} Introduction} %

  In non-relativistic quantum mechanics, a step potential which rises abruptly to a finite value at the origin and keeps its value to infinity plays the same role as a rigid wall in classical mechanics, because the reflection probability of the stationary scattering problem for the step becomes one when the incident energy is less than the height of the step. This is because the wave function under the step is essentially a real function and the corresponding probability current vanishes. This situation changes dramatically when one considers relativistic cases. Because the relativistic dispersion relation between energy and momentum allows solutions with, not only positive, but also negative energy, an oscillating solution can exist under the step if the value of the incident energy falls in a particular range called the Klein region, named after a paradoxical phenomenon for a relativistic electron described by the Dirac equation \cite{Klein1929}. Since the stationary solutions are understood to represent stationary flows of probability currents, the phenomenon implies the existence of a non-vanishing probability current under the step, which may be interpreted as a transmitted or tunneling current.

  Even though the implication seems physically counter-intuitive and, at the same time, suggestive when one combines it with the idea of Dirac sea, which may be lifted to produce electron-positron pairs when the applied potential is strong enough, it has generally been believed that the resolution of this problem could be sought only within the framework of quantum field theory. This is because the notion of particles and anti-particles is defined in quantum field theory, while in a single-particle quantum mechanics, one can just deal with positive- and negative-energy solutions with no clear association to particle and anti-particle. Even if the lack of the negative-energy solution might be interpreted as a presence of anti-particle, such an interpretation is only justified on the basis of quantum field theory.

  In quantum field theory, we calculate the scattering (S) matrix which describes the transition between two asymptotic states prepared at remote past $t = -\infty$ and future $t = \infty$ in terms of field operators and therefore it is crucial that one can define asymptotic states that allow a particle picture. When the potential is localized in space, the last condition seems to be satisfied on physical ground because particles are present only in regions far away from the potential at $t = \pm \infty$. This, on the contrary, means that for a potential that does not vanish at spatial infinity, like the step potential, it is not clear whether one can properly define an S matrix like in the usual cases, even if the potential is supposed to be switched on and off adiabatically.

  One may argue that such a difficulty is mainly due to a mathematical idealization and over-simplification of a physical setting, however, at the same time, the problem has attracted researchers for almost ninety years because its resolution is not only sought from an academic interest but also is hoped to bring us with some insight into the physical mechanism of pair creation of particle and anti-particle from the vacuum. In order to somehow bypass the difficulty, Nikishov proposed to interchange the role of time and spatial coordinates and to utilize stationary scattering solutions that have a single plane wave in the spatially asymptotic regions to construct the Green function \cite{Nikishov1969}. A similar idea is introduced to quantize the field variable of the scattering problem off the step potential, where solutions of the Dirac equation that satisfy a particular boundary condition, i.e., those composed of a single plane wave in regions outside of or under the step, are used to introduce creation/annihilation operators to discuss vacuum decay and pair creations \cite{Hansen1981, Calogeracos1999, Nikishov2004, Gavrilov2016, Chervyakov2018}. Notice that the dynamics, that is, the time development of the scattering process under a somewhat smeared potential (the Sauter potential \cite{Sauter1931}) has been analyzed numerically within the framework of quantum field theory \cite{Krekora2004}, where the field operator initially expanded in terms of solutions of the free Dirac equation is numerically simulated to see the effect of an incident fermion on the pair creation due to the Pauli exclusion principle. See also \cite{Cheng2010} for an introductory review.

  In this paper, we shall exclusively consider a relativistic fermion incident to a step potential with its energy lying in the Klein region and examine the scattering process within the framework of quantum field theory, but from a different point of view on the asymptotic states. The strategy adopted here may be considered more straightforward and to follow a naive physical expectation. After a brief review on the so-called Klein tunneling in Sec.\ \ref{Sec2} and the relevant stationary scattering solutions of the Dirac equation with the usual boundary condition in Sec.\ \ref{Sec3}, we introduce a field variable and expand it in terms of the stationary scattering solutions in Sec.\ \ref{Sec4}. It should be stressed that the fact that they form a complete orthonormal set is directly shown \cite{Ochiai2018}, so that the operators introduced as the expansion coefficients of the field variable are guaranteed to satisfy the standard anti-commutation relations. Then we examine the asymptotic limits $t \to \pm \infty$ of the field operator and find that there are four non-trivial limits existing, by which the asymptotic ``in'' and ``out'' operators are defined. It is shown that they are related through a similarity transformation, i.e., a Bogoliubov-like transformation. These asymptotic operators are used to define two asymptotic vacua in Sec.\ \ref{Sec5}. Their overlap can be explicitly evaluated to yield a vacuum decay constant. Under such an unstable vacuum, an incident particle with energy in the Klein region is shown to be completely reflected by the step, accompanying pairs of particle and anti-particle produced from the unstable vacuum in Sec.\ \ref{Sec6}. The final section \ref{Sec7} is devoted to a summary and discussions and Appendices \ref{AppendA} and \ref{AppendB} are added to fix the notation and to supply additional details and information.

\section{\label{Sec2} Klein tunneling in relativistic quantum mechanics} %

\begin{figure}[htbp]\centering 
    \includegraphics[width=0.7\columnwidth]{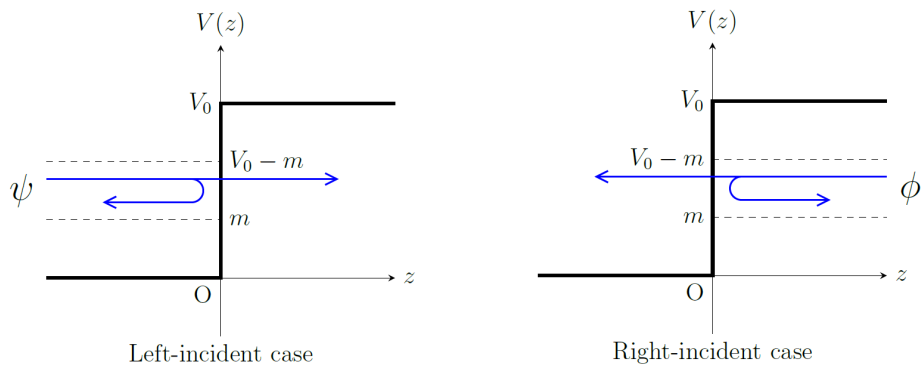} 
    \caption{Scattering states are categorized according to whether an incident flux is coming from the left ($\psi$) or right ($\phi$) of the step potential and to the range of their energies. } 
    \label{fig1:step_potential}%
  \end{figure}

  A scattering process of a Dirac fermion with mass $m$ by a one-dimensional step potential 
  \begin{align} 
    V(z) = \theta (z) V_{0}, \quad V_{0} > 2m, 
  \end{align} 
  is described by the Dirac Hamiltonian ($\hbar = c = 1$) 
  \begin{align} 
    H = H_{0} + V(z) = -i\alpha_{z} \partial_{z} + \beta m + V(z) 
    \label{eq:Hamiltonian(Sec2)}%
  \end{align} 
  and its stationary solution is easily obtained. Here $\theta (z)$ is the Heaviside step function and $\alpha_{z}$ and $\beta$ are two of the Dirac matrices. See Appendix \ref{AppendA} for the notation adopted here. We are mainly interested in the stationary scattering problem when the incident energy $E$ lies in the so-called Klein region, $m < E < V_{0} - m$. The stationary solution which is nothing but an eigenfunction of $H$ belonging to the energy eigenvalue $E$ is fixed once the boundary conditions and a continuity condition at $z = 0$ have been imposed on the eigenfunction (see Fig.\ \ref{fig1:step_potential}). We shall here treat the problem as an essentially one-dimensional scattering problem from the beginning and the trivial dependence on the other coordinates $x$ and $y$ will be ignored completely. 

  Within the framework of (single-particle) relativistic quantum mechanics, when a particle with momentum $p > 0$ and spin $s$ is incident from the left of the step, the stationary solution $\psi_{s}^{(E)} (z, t)$ with energy $E = \sqrt{p^{2} + m^{2}} \equiv E_{p}$ is expressed, in terms of the positive frequency solution $u(\pm p, s) e^{-iE_{p} t \pm ipz}$ for $z < 0$ (left of the step) and the negative frequency solution $v(q, s) e^{iE_{q} t - iqz}$ for $z > 0$ (under the step), as ($E = E_{p} = V_{0} - E_{q}, \, E_{q} = \sqrt{q^{2} + m^{2}}, \, q > 0$) 
  \begin{align} 
    \psi_{s}^{(E)} (z, t) = \frac{1}{\sqrt{2\pi}} \sqrt{\frac{m}{E}} \left[ \theta (-z) \left\{ u(p, s) e^{ipz} + Ru(-p, s) e^{-ipz} \right\} + \theta (z) T_{\to} \sigma_{z} v(q, s) e^{-iqz} \right] e^{-iEt}, 
    \label{eq:leftincident_solution(Sec2)}%
  \end{align} 
  with 
  \begin{align} 
    R = \frac{\sqrt{\frac{V_{0} - E - m}{E + m}} - \sqrt{\frac{V_{0} - E + m}{E - m}}}{\sqrt{\frac{V_{0} - E - m}{E + m}} + \sqrt{\frac{V_{0} - E + m}{E - m}}}, \quad T_{\to} = \frac{2}{\sqrt{\frac{V_{0} - E - m}{E + m}} + \sqrt{\frac{V_{0} - E + m}{E - m}}}. 
  \end{align} 
  (Appendix \ref{AppendA} summarizes the details of spinors.) The continuity of the current defined by $j_{z} = \bar{\psi} \gamma^{3} \psi = \psi^{\dagger} \alpha_{z} \psi$ at $z = 0$ implies the consevation of probability 
  \begin{align} 
    \frac{p}{m} - |R|^{2} \frac{p}{m} = |T_{\to}|^{2} \frac{q}{m} \quad \longrightarrow \quad P_{r} + P_{t} = |R|^{2} + \frac{q}{p} |T_{\to}|^{2} = 1. 
    \label{eq:current_conservation(Sec2)}%
  \end{align} 
  Recall that this is a normal conservation law of probability, but a finite and non-vanishing transmission probability, which does not vanish even at the infinite-potential limit $V_{0} \to \infty$, survives 
  \begin{align} 
    P_{t} = \frac{q}{p} |T_{\to}|^{2} \quad \xrightarrow{V_{0} \to \infty} \quad \frac{2p}{E_{p} + p} = \frac{2\sqrt{E^{2} - m^{2}}}{E + \sqrt{E^{2} - m^{2}}} \neq 0. 
  \end{align} 
  This phenomenon is known as the Klein tunneling \cite{Calogeracos1999}.

  Owing to the presence of negative energy solutions to the Dirac equation, we can set up, not only the left-incident scattering problem, but also the right-incident problem, because even if the energy is below the potential height, there exist oscillating solutions of the equation. In fact, we obtain the right-incident solution $\phi_{s}^{(E)} (z, t)$ with a negative incident momentum $-q < 0$ and spin $s$ ($E = V_{0} - E_{q} = E_{p}, \, -p < 0$) 
  \begin{align} 
    \phi_{s}^{(E)} (z, t) = \frac{1}{\sqrt{2\pi}} \sqrt{\frac{m}{V_{0} - E}} \left[ \theta (-z) T_{\gets} \sigma_{z} u(-p, s) e^{-ipz} + \theta (z) \left\{ v(-q, s) e^{iqz} + Rv(q, s) e^{-iqz} \right\} \right] e^{-iEt}, 
  \end{align} 
  where 
  \begin{align} 
    R = -\frac{\sqrt{\frac{E + m}{V_{0} - E - m}} - \sqrt{\frac{E - m}{V_{0} - E + m}}}{\sqrt{\frac{E + m}{V_{0} - E - m}} + \sqrt{\frac{E - m}{V_{0} - E + m}}}, \quad T_{\gets} = -\frac{2}{\sqrt{\frac{E + m}{V_{0} - E - m}} + \sqrt{\frac{E - m}{V_{0} - E + m}}}. 
  \end{align} 
  The continuity of the current results in the conservation of probability 
  \begin{align} 
    -\frac{p}{m} |T_{\gets}|^{2} = -\frac{q}{m} + \frac{q}{m} |R|^{2}, 
  \end{align} 
  which is essentially the same as \eqref{eq:current_conservation(Sec2)}. The reflection coefficient $R$ turns out to be the same as in the left-incident case, which is nothing but a realization of reciprocity in quantum mechanics.

  It would be worth mentioning that a choice of $v(-q, s) e^{iqz}$, instead of $v(q, s) e^{-iqz}$, in \eqref{eq:leftincident_solution(Sec2)} results in a negative factor in front of the transmission coefficient squared in \eqref{eq:current_conservation(Sec2)} and the reflection probability becomes greater than one, which had been known as the Klein paradox \cite{Calogeracos1999}. It is to be observed that in such a case, the sign of the current (and also the group velocity) in the potential region ($z > 0$) becomes negative, which is not considered to satisfy the boundary condition for the left-incident scattering problem where only a positive current (transmitted current) is admissible for $z > 0$.

\section{\label{Sec3} Scattering states} %

  The stationary solutions for the left- and right-incident scattering problems are given by the eigenstates of the Hamiltonian $H$ \eqref{eq:Hamiltonian(Sec2)} and are characterized by their boundary condition, i.e., left-incident ($\psi$) or right-incident ($\phi$), and their eigenvalue $E$. These eigenfunctions form a complete orthonormal set. Here only those cases where the potential step is higher than $2m$, $V_{0} > 2m$, are considered.

  First, the orthogonality of the scattering states follows from the general argument for hermitian Hamiltonians, that is, eigenfunctions belonging to different eigenvalues are orthogonal to each other. Two $\psi$'s with different energies are orthogonal and normalized as 
  \begin{align} 
    \int_{-\infty}^{\infty} dz \psi_{s}^{(E) \dagger} (z) \psi_{s'}^{(E')} (z) = \theta (EE') \delta (p - p') \delta_{s, s'}, 
    \label{eq:orthonormality_psi(Sec3)}%
  \end{align} 
  where $|E| = E_{p}, \, |E'| = E_{p'}$. Similarly, we should have 
  \begin{align} 
    \int_{-\infty}^{\infty} dz \phi_{s}^{(E) \dagger} (z) \phi_{s'}^{(E')} (z) = \theta \left[ (E - V_{0})(E' - V_{0}) \right] \delta (q - q') \delta_{s, s'}, 
    \label{eq:orthonormality_phi(Sec3)}%
  \end{align} 
  where $|E - V_{0}| = E_{q}, \, |E' - V_{0}| = E_{q'}$, and 
  \begin{align} 
    \int_{-\infty}^{\infty} dz \psi_{s}^{(E) \dagger} (z) \phi_{s'}^{(E')} (z) = 0. 
    \label{eq:orthogonality(Sec3)}%
  \end{align} 
  Since the last orthogonality relation does not follow from the general argument for both $\psi_{s}^{(E)}$ and $\phi_{s'}^{(E')}$ can happen to belong to the same energy $E = E'$, its validity has to be examined separately (see Appendix \ref{AppendB}). The orthonormality conditions imply that the following form of completeness relation holds 
  \begin{align} 
    \sum_{s, r = \pm} \int_{0}^{\infty} dp \psi_{s}^{(rE_{p})} (z) \psi_{s}^{(rE_{p}) \dagger} (z') + \sum_{s, r = \pm} \int_{0}^{\infty} dq \phi_{s}^{(V_{0} + rE_{q})} (z) \phi_{s}^{(V_{0} + rE_{q}) \dagger} (z') = \delta (z - z') \1_{4\times 4}, 
    \label{eq:completeness(Sec3)}%
  \end{align} 
  where $\1_{4\times 4}$ is the unit matrix acting in the spinor space. Notice that the fact that they actually constitute a complete orthonormal set is shown in a straightforward way, i.e., the relation \eqref{eq:completeness(Sec3)} has been shown explicitly in \cite{Ochiai2018}, though the proof itself first appeared more than forty years ago \cite{Ruijsenaars1977}.

\section{\label{Sec4} Quantized field} %

  In order to discuss the scattering process under the Hamiltonian \eqref{eq:Hamiltonian(Sec2)} within the framework of quantum field theory, we introduce a field operator $\Psi (z, t)$ and set up the equal-time anti-commutation relations between $\Psi$s and $\Psi^{\dagger}$s, only non-vanishing ones of which are 
  \begin{align} 
    \{ \Psi_{\alpha} (z, t), \, \Psi_{\beta}^{\dagger} (z', t) \} = \delta_{\alpha, \beta} \delta (z - z'). 
    \label{eq:ACR(Sec4)}%
  \end{align} 
  The field operator $\Psi$ is then expanded in terms of a complete orthonormal set and creation and annihilation operators are introduced as the expansion coefficients. Since we are interested in the scattering process described by the Hamiltonian \eqref{eq:Hamiltonian(Sec2)}, it would be natural to choose its eigenfunctions that satisfy the boundary conditions for scattering processes, that is, the left-incident ($\psi$) and right-incident ($\phi$) scattering states shall be chosen as the basis functions.

  We introduce operators $b_{l (r)}, \, d_{l (r)}^{\dagger}$ as expansion coefficients of $\Psi$ expanded in terms of the scattering states $\psi (\phi)$ 
  \begin{equation}\begin{split} 
    \Psi (z, t) &= \int_{0}^{\infty} dp \sum_{s} \left[ b_{l} (p, s) \psi_{s}^{(E_{p})} (z, t) + d_{l}^{\dagger} (p, s) \psi_{s}^{(-E_{p})} (z, t) \right] \\ 
    &\quad + \int_{0}^{\infty} dq \sum_{s} \left[ b_{r} (q, s) \phi_{s}^{(V_{0} + E_{q})} (z, t) + d_{r}^{\dagger} (q, s) \phi_{s}^{(V_{0} - E_{q})} (z, t) \right]. 
  \end{split}\end{equation} 
  The orthonormality of scattering eigenstates $\psi_{s}^{(E)}$ and $\phi_{s}^{(E)}$ guarantees the usual canonical anti-commutation relations between creation/annihilation operators. Indeed operators are expressed as 
  \begin{align} 
    b_{l} (p, s) &= \int_{-\infty}^{\infty} dz \psi_{s}^{(E_{p}) \dagger} (z, t) \Psi (z, t), & b_{r} (q, s) &= \int_{-\infty}^{\infty} dz \phi_{s}^{(V_{0} + E_{q}) \dagger} (z, t) \Psi (z, t), \\ 
    d_{l}^{\dagger} (p, s) &= \int_{-\infty}^{\infty} dz \psi_{s}^{(-E_{p}) \dagger} (z, t) \Psi (z, t), & d_{r}^{\dagger} (q, s) &= \int_{-\infty}^{\infty} dz \phi_{s}^{(V_{0} - E_{q}) \dagger} (z, t) \Psi (z, t) 
  \end{align} 
  and since the right-hand sides becomes all time-independent if $\Psi$ satisfies the same equation of motion as $\psi_{s}^{(E)}$ and $\phi_{s}^{(E)}$, the equal-time anti-commutation relations \eqref{eq:ACR(Sec4)} and the orthonormality of $\psi_{s}^{(E)}$ and $\phi_{s}^{(E)}$ \eqref{eq:orthonormality_psi(Sec3)}--\eqref{eq:orthogonality(Sec3)} are enough to show that the operators satisfy the standard anti-commutation relations.

  Since the potential extending to $z = \infty$ is present at all times, it is not clear whether the standard construction of asymptotic states at $t = \pm \infty$ is consistently applied, especially when one considers the scattering process occurring in the Klein energy range. We can think, however, that the field operator $\Psi$ contains all information of scattering and endeavor to extract asymptotic information by considering appropriate limits. We consider exclusively the case where the Klein tunneling occurs, i.e., the left-incident energy $E > m$ is below the potential height, $E = E_{p} < V_{0} - m$. Consider, for example, the following limit 
  \begin{align} 
    \lim_{t \to -\infty} \int_{-\infty}^{\infty} dz u_{p, s}^{\dagger} (z, t) \Psi (z, t), 
    \label{eq:limit_innerprod(Sec4)}%
  \end{align} 
  where 
  \begin{align} 
    u_{p, s} (z, t) = \frac{1}{\sqrt{2\pi}} \sqrt{\frac{m}{E_{p}}} u(p, s) e^{ipz - iE_{p} t} \quad (p > 0) 
  \end{align} 
  is nothing but the (normalized) positive-energy solution of the free Dirac equation. The integration over $z$ in \eqref{eq:limit_innerprod(Sec4)} results in one of the following forms, for those terms with energy lying in the range $m < E < V_{0} - m$, 
  \begin{align} 
    &\int_{-\infty}^{0} dz e^{-ipz + iE_{p} t} e^{ip'z - iE_{p'} t} = \frac{-i}{p' - p - i\epsilon} e^{-i(E_{p'} - E_{p}) t} \rightarrow 2\pi \delta (p' - p), \\ 
    &\int_{-\infty}^{0} dz e^{-ipz + iE_{p} t} e^{-ip'z - iE_{p'} t} = \frac{i}{p' + p} e^{-i(E_{p'} - E_{p}) t} \rightarrow 0, \\ 
    &\int_{0}^{\infty} dz e^{-ipz + iE_{p} t} e^{\pm iq'z - i(V_{0} - E_{q'}) t} = \frac{-i}{\pm q' - p} e^{-i(V_{0} - E_{q'} - E_{p}) t} \rightarrow 0, 
  \end{align} 
  in the $t \to -\infty$ limit, owing to the Riemann--Lebesgue lemma. The lemma also implies that the other terms in \eqref{eq:limit_innerprod(Sec4)} all disappear in the limit because their energy does not match $E_{p}$ and oscillates indefinitely. The limit \eqref{eq:limit_innerprod(Sec4)} can be interpreted as the asymptotic annihilation operator for in-coming particle $b_{\mathrm{in}} (p, s)$ associated with the positive-energy solution and we find 
  \begin{align} 
    b_{\mathrm{in}} (p, s) = b_{l} (p, s). 
  \end{align} 
  Observe that the above limit \eqref{eq:limit_innerprod(Sec4)} just extracts the coefficient of $u_{p, s} (z, t)$ in $\Psi$. The field operator $\Psi$ also contains other solutions of the Dirac equation, $u_{-p, s} (z, t)$ and 
  \begin{align} 
    v_{\pm q, s} (z, t) = \frac{1}{\sqrt{2\pi}} \sqrt{\frac{m}{E_{q}}} v(\pm q, s) e^{\mp iqz - i(V_{0} - E_{q}) t} \quad (q > 0), 
  \end{align} 
  through the scattering wave functions $\psi$ and $\phi$ in the Klein energy range. It would be natural to define the asymptotic creation operator for in-coming anti-particle $d_{\mathrm{in}}^{\dagger} (q, s)$ by the limit 
  \begin{align} 
    d_{\mathrm{in}}^{\dagger} (q, s) = \lim_{t \to -\infty} \int_{-\infty}^{\infty} dz v_{-q, s}^{\dagger} (z, t) \Psi (z, t), 
    \label{eq:def_din(Sec4)}%
  \end{align} 
  which results in the relation 
  \begin{align} 
    d_{\mathrm{in}}^{\dagger} (q, s) = d_{r}^{\dagger} (q, s). 
  \end{align} 
  We understand that the limits \eqref{eq:limit_innerprod(Sec4)} and \eqref{eq:def_din(Sec4)} but with opposite momenta $p \to -p$ and $-q \to q$ give nothing 
  \begin{align} 
    \lim_{t \to -\infty} \int_{-\infty}^{\infty} dz u_{-p, s}^{\dagger} (z, t) \Psi (z, t) = \lim_{t \to -\infty} \int_{-\infty}^{\infty} dz v_{q, s}^{\dagger} (z, t) \Psi (z, t) = 0, 
  \end{align} 
  as expected from the directions of in-coming particle and anti-particle in the Klein region. These quantities become non-vanishing, instead in the other asymptotic limit, i.e., at $t = \infty$, and they shall be denoted as $b_{\mathrm{out}}$ and $d_{\mathrm{out}}^{\dagger}$, i.e., the particle--anti-particle interpretation for out-going states. We evaluate the limit to obtain, for example, 
  \begin{align} 
    b_{\mathrm{out}} (p, s) = \lim_{t \to \infty} \int_{-\infty}^{\infty} dz u_{-p, s}^{\dagger} (z, t) \Psi (z, t) = R(p) b_{l} (p, s) - \sqrt{\frac{E_{q}}{E_{p}}} T(p) d_{r}^{\dagger} (q, \tilde{s}), 
    \label{eq:def_bout(Sec4)}%
  \end{align} 
  where $E_{p} = V_{0} - E_{q}$, $R$ and $T = T_{\to}$ are reflection and transmission coefficients appearing in the left-incident scattering eigenfunction $\psi$ and the spin $\tilde{s}$ stands for the spinor $\tilde{\xi} (\tilde{s}) = \sigma_{z} \xi (s)$. In deriving \eqref{eq:def_bout(Sec4)}, the following relations 
  \begin{align} 
    \delta (p - p') = \frac{p}{E_{p}} \frac{E_{q}}{q} \delta (q - q'), \quad T_{\gets} (q) = -\frac{q}{p} T_{\to} (p) 
  \end{align} 
  have been used. It is interesting to see that this operator $b_{\mathrm{out}}$ can be expressed as a similarity transformation of $b_{l}$ multiplied by $R$ 
  \begin{align} 
    b_{\mathrm{out}} (p, s) = R(p) e^{B^{\dagger}} b_{l} (p, s) e^{-B^{\dagger}}, \quad B^{\dagger} = \sum_{s} \int \limits_{E_{p} < V_{0} - m} dp \sqrt{\frac{E_{q}}{E_{p}}} \frac{T(p)}{R(p)} b_{l}^{\dagger} (p, s) d_{r}^{\dagger} (q, \tilde{s}) 
    \label{eq:trans_bout(Sec4)}%
  \end{align} 
  and satisfies, together with its hermitian conjugate, the standard anti-commutation relation 
  \begin{equation}\begin{split} 
    \{ b_{\mathrm{out}} (p, s), \, b_{\mathrm{out}}^{\dagger} (p', s') \} &= R^{2} (p) \delta (p - p') \delta_{s, s'} + \frac{E_{q}}{E_{p}} T^{2} (p) \delta (q - q') \delta_{\tilde{s}, \tilde{s}'} \\
    &= \left[ R^{2} (p) + \frac{q}{p} T^{2} (p) \right] \delta (p - p') \delta_{s, s'} = \delta (p - p') \delta_{s, s'}, 
  \end{split}\end{equation} 
  where the last equality follows from the fact that the quantity in the square parentheses is unity, which is nothing but the current conservation condition. Another limit at $t = \infty$ worth evaluating corresponds to the anti-particle creation operator $d_{\mathrm{out}}^{\dagger}$ and yields 
  \begin{align} 
    d_{\mathrm{out}}^{\dagger} (q, s) = \lim_{t \to \infty} \int_{-\infty}^{\infty} dz v_{q, s}^{\dagger} (z, t) \Psi (z, t) = R(p) d_{r}^{\dagger} (q, s) + \frac{q}{p} \sqrt{\frac{E_{p}}{E_{q}}} T(p) b_{l} (p, \tilde{s}), 
    \label{eq:def_dout(Sec4)}%
  \end{align} 
  where use has made of $R(p) \equiv R_{\gets} (p) = R_{\to} (q)$ (the reciprocity). This operator is written as 
  \begin{align} 
    d_{\mathrm{out}}^{\dagger} (q, s) = R(p) e^{-C} d_{r}^{\dagger} (q, s) e^{C}, 
  \end{align} 
  where the operator $C$ turns out to be the same as $B$ 
  \begin{align} 
    C = \sum_{s} \int \limits_{E_{q} < V_{0} - m} dq \frac{q}{p} \sqrt{\frac{E_{p}}{E_{q}}} \frac{T(p)}{R(p)} d_{r} (q, s) b_{l} (p, \tilde{s}) = \sum_{s} \int \limits_{E_{p} < V_{0} - m} dp \sqrt{\frac{E_{q}}{E_{p}}} \frac{T(p)}{R(p)} d_{r} (q, \tilde{s}) b_{l} (p, s) = B, 
  \end{align} 
  and satisfies 
  \begin{align} 
    \{ d_{\mathrm{out}} (q, s), \, d_{\mathrm{out}}^{\dagger} (q', s') \} = R^{2} (p) \delta (q - q') \delta_{s, s'} + \left( \frac{q}{p} \right)^{2} \frac{E_{p}}{E_{q}} T^{2} (p) \delta (p - p') \delta_{\tilde{s}, \tilde{s}'} = \delta (q - q') \delta_{s, s'}. 
  \end{align}

\section{\label{Sec5} Asymptotic vacua} %

  We are now in a position to define asymptotic vacuum states. We define an asymptotic vacuum state at $t = -\infty$, called ``in'' vacuum, as the state that is annihilated by the asymptotic operators $b_{\mathrm{in}}$ and $d_{\mathrm{in}}$ 
  \begin{align} 
    b_{\mathrm{in}} (p, s) \ket{0}_{\mathrm{in}} = d_{\mathrm{in}} (q, s) \ket{0}_{\mathrm{in}} = 0. 
  \end{align} 
  The ``in'' vacuum $\ket{0}_{\mathrm{in}}$ is nothing but the normalized state $\ket{0}$ that is annihilated by the operators $b_{l}$ and $d_{r}$  
  \begin{align} 
    \ket{0}_{\mathrm{in}} = \ket{0}, \quad b_{l} (p, s) \ket{0} = d_{r} (q, s) \ket{0} = 0, \quad \braket{0|0} = 1. 
  \end{align} 
  We also introduce another asymptotic vacuum state at $t = \infty$, called ``out'' vacuum. They shall be defined as the state that is annihilated by the operators $b_{\mathrm{out}}$ and $d_{\mathrm{out}}$ 
  \begin{align} 
    b_{\mathrm{out}} (p, s) \ket{0}_{\mathrm{out}} &= R(p) \left( b_{l} (p, s) - \sqrt{\frac{E_{q}}{E_{p}}} \frac{T(p)}{R(p)} d_{r}^{\dagger} (q, \tilde{s}) \right) \ket{0}_{\mathrm{out}} = 0, \\ 
    d_{\mathrm{out}} (q, \tilde{s}) \ket{0}_{\mathrm{out}} &= R(p) \left( d_{r} (q, s) + \frac{q}{p} \sqrt{\frac{E_{p}}{E_{q}}} \frac{T(p)}{R(p)} b_{l}^{\dagger} (p, \tilde{s}) \right) \ket{0}_{\mathrm{out}} = 0. 
  \end{align} 
  It is not difficult to see that the ``out'' vacuum is explicitly constructed as 
  \begin{align} 
    \ket{0}_{\mathrm{out}} = \prod_{p, s} R(p) \left( 1 + \frac{2\pi}{L} \sqrt{\frac{E_{q}}{E_{p}}} \frac{T(p)}{R(p)} b_{l}^{\dagger} (p, s) d_{r}^{\dagger} (q, \tilde{s}) \right) \ket{0}, 
    \label{eq:outvac_writtenby_invac(Sec5)}%
  \end{align} 
  where $L$ is the (infinite) size of the physical system and is formally equal to $2\pi \delta (p - p)$. To see that it is annihilated by $b_{\mathrm{out}} (p, s)$ and $d_{\mathrm{out}} (q, \tilde{s})$, we only need to confirm that $\forall p > 0$ ($E_{p} = V_{0} - E_{q}$) and $\forall s$, the relations 
  \begin{align} 
    \left( b_{l} (p, s) - \sqrt{\frac{E_{q}}{E_{p}}} \frac{T(p)}{R(p)} d_{r}^{\dagger} (q, \tilde{s}) \right) \left( 1 + \frac{2\pi}{L} \sqrt{\frac{E_{q}}{E_{p}}} \frac{T(p)}{R(p)} b_{l}^{\dagger} (p, s) d_{r}^{\dagger} (q, \tilde{s}) \right) \ket{0} &= 0, \\ 
    \left( d_{r} (q, s) + \frac{q}{p} \sqrt{\frac{E_{p}}{E_{q}}} \frac{T(p)}{R(p)} b_{l}^{\dagger} (p, \tilde{s}) \right) \left( 1 + \frac{2\pi}{L} \sqrt{\frac{E_{q}}{E_{p}}} \frac{T(p)}{R(p)} b_{l}^{\dagger} (p, s) d_{r}^{\dagger} (q, \tilde{s}) \right) \ket{0} &= 0 
  \end{align} 
  hold. The ``out'' vacuum $\ket{0}_{\mathrm{out}}$ is normalized to unity, as can be shown explicitly 
  \begin{equation}\begin{split} 
    {}_{\mathrm{out}} \! \braket{0|0}_{\mathrm{out}} &= \bra{0} \prod_{p, s} R^{2} (p) \left( 1 + \frac{2\pi}{L} \sqrt{\frac{E_{q}}{E_{p}}} \frac{T(p)}{R(p)} d_{r} (q, \tilde{s}) b_{l} (p, s) \right) \left( 1 + \frac{2\pi}{L} \sqrt{\frac{E_{q}}{E_{p}}} \frac{T(p)}{R(p)} b_{l}^{\dagger} (p, s) d_{r}^{\dagger} (q, \tilde{s}) \right) \ket{0} \\ 
    &= \prod_{p, s} R^{2} (p) \left( 1 + \frac{q}{p} \frac{T^{2} (p)}{R^{2} (p)} \right) = 1, 
  \end{split}\end{equation} 
  where the last equality follows from the current conservation condition $R^{2} (p) + \frac{q}{p} T^{2} (p) = 1$. It may be interesting to see that the ``out'' vacuum state can be expressed as $\ket{0}_{\mathrm{out}} = \mathcal{N} e^{B^{\dagger}} \ket{0}$, where the operator $B^{\dagger}$ is the same as in \eqref{eq:trans_bout(Sec4)} and the normalization constant is formally given by $\mathcal{N} = \prod_{p, s} R(p)$. Notice that the meaning of an infinite product over continuous variable $p$ is unclear and its precise meaning shall be given in the following.

  We are interested in the transition amplitudes to find specific asymptotic states at $t = \infty$ when the initial state is prepared at $t = -\infty$. It is thus convenient to express ``in'' states in terms of ``out'' states. We can, for example, invert the relation \eqref{eq:def_bout(Sec4)} and \eqref{eq:def_dout(Sec4)} to express ``in'' operators in terms of ``out'' operators 
  \begin{align} 
    \begin{pmatrix} b_{\mathrm{in}} (p, s) \\ d_{\mathrm{in}}^{\dagger} (q, \tilde{s}) \end{pmatrix} = \begin{pmatrix} R(p) & \sqrt{\frac{E_{q}}{E_{p}}} T(p) \\ -\frac{q}{p} \sqrt{\frac{E_{p}}{E_{q}}} T(p) & R(p) \end{pmatrix} \begin{pmatrix} b_{\mathrm{out}} (p, s) \\ d_{\mathrm{out}}^{\dagger} (q, \tilde{s}) \end{pmatrix} 
  \end{align} 
  and we also have, instead of \eqref{eq:outvac_writtenby_invac(Sec5)}, 
  \begin{align} 
    \ket{0}_{\mathrm{in}} = \prod_{p, s} R^{-1} (p) \left( 1 - \frac{2\pi}{L} \sqrt{\frac{E_{q}}{E_{p}}} \frac{T(p)}{R(p)} b_{\mathrm{in}}^{\dagger} (p, s) d_{\mathrm{in}}^{\dagger} (q, \tilde{s}) \right) \ket{0}_{\mathrm{out}}. 
  \end{align} 
  It is not difficult to see that the last expression is actually symmetric between ``in'' and ``out'' for it is rewritten, in terms of ``out'' operators, as 
  \begin{align} 
    \ket{0}_{\mathrm{in}} = \prod_{p, s} R (p) \left( 1 - \frac{2\pi}{L} \sqrt{\frac{E_{q}}{E_{p}}} \frac{T(p)}{R(p)} b_{\mathrm{out}}^{\dagger} (p, s) d_{\mathrm{out}}^{\dagger} (q, \tilde{s}) \right) \ket{0}_{\mathrm{out}}. 
    \label{eq:invac_writtenby_outvac(Sec5)}%
  \end{align}

  Notice that the expression \eqref{eq:invac_writtenby_outvac(Sec5)}, which is formal and somewhat ambiguous, can be written as 
  \begin{align} 
    \ket{0}_{\mathrm{in}} = \mathcal{N} e^{-B_{\mathrm{out}}^{\dagger}} \ket{0}_{\mathrm{out}}, 
  \end{align} 
  where 
  \begin{align} 
    B_{\mathrm{out}}^{\dagger} = \sum_{s} \int \limits_{E_{p} < V_{0} - m} dp \sqrt{\frac{E_{q}}{E_{p}}} \frac{T(p)}{R(p)} b_{\mathrm{out}}^{\dagger} (p, s) d_{\mathrm{out}}^{\dagger} (q, \tilde{s}). 
    \label{eq:Bout(Sec5)}%
  \end{align} 
  It is easily seen that the state $e^{-B_{\mathrm{out}}^{\dagger}} \ket{0}_{\mathrm{out}}$ is actually annihilated by $b_{\mathrm{in}}$ and $d_{\mathrm{in}}$ 
  \begin{align} 
    b_{\mathrm{in}} (p, s) \ket{0}_{\mathrm{in}} &\propto \left( R(p) b_{\mathrm{out}} (p, s) + \sqrt{\frac{E_{q}}{E_{p}}} T(p) d_{\mathrm{out}}^{\dagger} (q, \tilde{s}) \right) e^{-B_{\mathrm{out}}^{\dagger}} \ket{0}_{\mathrm{out}} = e^{-B_{\mathrm{out}}^{\dagger}} R(p) b_{\mathrm{out}} (p, s) \ket{0}_{\mathrm{out}} = 0, \\ 
    d_{\mathrm{in}} (q, \tilde{s}) \ket{0}_{\mathrm{in}} &\propto \left( R(p) d_{\mathrm{out}} (q, \tilde{s}) - \frac{q}{p} \sqrt{\frac{E_{p}}{E_{q}}} T(p) b_{\mathrm{out}}^{\dagger} (p, s) \right) e^{-B_{\mathrm{out}}^{\dagger}} \ket{0}_{\mathrm{out}} = e^{-B_{\mathrm{out}}^{\dagger}} R(p) d_{\mathrm{out}} (q, \tilde{s}) \ket{0}_{\mathrm{out}} = 0. 
  \end{align} 
  The constant $\mathcal{N}$ is calculated from the normalization condition $\mathcal{N}^{-2} = {}_{\mathrm{out}} \! \braket{0 | e^{-B_{\mathrm{out}}} e^{-B_{\mathrm{out}} \dagger} | 0}_{\mathrm{out}}$.

  In order to estimate $\mathcal{N}$, consider a functional 
  \begin{align} 
    F[a] \equiv \braket{0 | e^{-B[a]} e^{-B^{\dagger} [a]} | 0}, \quad B[a] = \sum_{s} \int \limits_{E_{p} < V_{0} - m} dp \, a(p, s) d(q, \tilde{s}) b(p, s), 
  \end{align} 
  where the index ``out'' has been suppressed for notational simplicity. A functional derivative of $F$ brings down operators $db$ and $b^{\dagger} d^{\dagger}$ 
  \begin{align} 
    \frac{\delta}{\delta a(p, s)} F[a] = -\braket{0 | e^{-B[a]} \left( d(q, \tilde{s}) b(p, s) + b^{\dagger} (p, s) d^{\dagger} (q, \tilde{s}) \right) e^{-B^{\dagger} [a]} | 0}. 
  \end{align} 
  These operators can be moved to, say, just the right of $\bra{0}$ 
  \begin{equation}\begin{split} 
    &-\bra{0} \left[ d(q, \tilde{s}) b(p, s) + \biggl( b^{\dagger} (p, s) - a(p, s) d(q, \tilde{s}) \biggr) \left( d^{\dagger} (q, \tilde{s}) + \frac{q}{p} \frac{E_{p}}{E_{q}} a(p, s) b(p, s) \right) \right] e^{-B[a]} e^{-B^{\dagger} [a]} \ket{0} \\ 
    &= a(p, s) \braket{0 | d(q, \tilde{s}) d^{\dagger} (q, \tilde{s}) | 0} \braket{0 | e^{-B[a]} e^{-B^{\dagger} [a]} | 0} - \braket{0 | d(q, \tilde{s}) b(p, s) e^{-B[a]} e^{-B^{\dagger} [a]} | 0} \left( 1 - \frac{q}{p} \frac{E_{p}}{E_{q}} a^{2} (p, s) \right). 
  \end{split}\end{equation} 
  The last matrix element satisfies the relations 
  \begin{equation}\begin{split} 
    &\braket{0 | d(q, \tilde{s}) b(p, s) e^{-B[a]} e^{-B^{\dagger} [a]} | 0} \\ 
    &= \bra{0} e^{-B[a]} e^{-B^{\dagger} [a]} \left( d(q, \tilde{s}) + \frac{q}{p} \frac{E_{p}}{E_{q}} a(p, s) b^{\dagger} (p, s) \right) \biggl( b(p, s) - a(p, s) d^{\dagger} (q, \tilde{s}) \biggr) \ket{0} \\ 
    &= -a(p, s) \braket{0 | e^{-B[a]} e^{-B^{\dagger} [a]} | 0} \braket{0 | d(q, \tilde{s}) d^{\dagger} (q, \tilde{s}) | 0} - \braket{0 | e^{-B[a]} e^{-B^{\dagger} [a]} b^{\dagger} (p, s) d^{\dagger} (q, \tilde{s}) | 0} \frac{q}{p} \frac{E_{p}}{E_{q}} a^{2} (p, s) \\ 
    &= -a(p, s) \left( 1 - \frac{q}{p} \frac{E_{p}}{E_{q}} a^{2} (p, s) \right) \braket{0 | e^{-B[a]} e^{-B^{\dagger} [a]} | 0} \braket{0 | d(q, \tilde{s}) d^{\dagger} (q, \tilde{s}) | 0} \\ 
    &\quad + \braket{0 | d(q, \tilde{s}) b(p, s) e^{-B[a]} e^{-B^{\dagger} [a]} | 0} \left( \frac{q}{p} \frac{E_{p}}{E_{q}} a^{2} (p, s) \right)^{2}, 
  \end{split}\end{equation} 
  which results in 
  \begin{align} 
    \braket{0 | d(q, \tilde{s}) b(p, s) e^{-B[a]} e^{-B^{\dagger} [a]} | 0} = -\frac{a(p, s)}{1 + \frac{q}{p} \frac{E_{p}}{E_{q}} a^{2} (p, s)} \braket{0 | e^{-B[a]} e^{-B^{\dagger} [a]} | 0} \braket{0 | d(q, \tilde{s}) d^{\dagger} (q, \tilde{s}) | 0}. 
  \end{align} 
  We thus understand that $F[a]$ satisfies 
  \begin{align} 
    \frac{\delta}{\delta a(p, s)} F[a] = \frac{2a(p, s)}{1 + \frac{q}{p} \frac{E_{p}}{E_{q}} a^{2} (p, s)} \braket{0 | d(q, \tilde{s}) d^{\dagger} (q, \tilde{s}) | 0} F[a] = \frac{2\frac{q}{p} \frac{E_{p}}{E_{q}} a(p, s)}{1 + \frac{q}{p} \frac{E_{p}}{E_{q}} a^{2} (p, s)} \braket{0 | b(p, s) b^{\dagger} (p, s) | 0} F[a], 
  \end{align} 
  the solution of which with the condition $F[0] = 1$ reads as 
  \begin{align} 
    F[a] = \exp \left[ \sum_{s} \int \limits_{E_{p} < V_{0} - m} dp \ln \left( 1 + \frac{q}{p} \frac{E_{p}}{E_{q}} a^{2} (p, s) \right) \braket{0 | b(p, s) b^{\dagger} (p, s) | 0} \right]. 
  \end{align} 
  This concludes that the normalization constant $\mathcal{N}$, which is reproduced by the functional $F[a]$ once $a(p, s)$ is replaced with $\sqrt{E_{q}/E_{p}} \, T(p)/R(p)$, is given by \cite{comment:vac-to-vac} 
  \begin{equation}\begin{split} 
    \mathcal{N} &= \exp \left[ -\frac{1}{2} \sum_{s} \int \limits_{E_{p} < V_{0} - m} dp \ln \left( 1 + \frac{q}{p} \frac{T^{2} (p)}{R^{2} (p)} \right) \braket{0 | b(p, s) b^{\dagger} (p, s) | 0} \right] \\ 
    &= \exp \left[ \frac{1}{2} \sum_{s} \int \limits_{E_{p} < V_{0} - m} dp \ln R^{2} (p) \braket{0 | b(p, s) b^{\dagger} (p, s) | 0} \right]. 
  \end{split}\end{equation}

  Since the matrix element in the exponent diverges $\braket{0 | b(p, s) b^{\dagger} (p, s) | 0} = \delta (p - p) = \infty$, this constant $\mathcal{N}$ formally vanishes, implying that the two vacua, $\ket{0}_{\mathrm{in}}$ and $\ket{0}_{\mathrm{out}}$, are orthogonal and have no overlaps, for $\mathcal{N} = {}_{\mathrm{out}} \! \braket{0|0}_{\mathrm{in}}$. Observe, however, that if we replace the delta function in momentum space with that in energy space, by $\delta (p - p) = \delta (E_{p} - E_{p}) \, p/E_{p}$, and recall that the last delta function is related to the total time $T_{\mathrm{total}}$ by $\delta (E_{p} - E_{p}) = T_{\mathrm{total}} / (2\pi)$, we may interpret that the ``in'' vacuum decays with a decay rate $\gamma$ 
  \begin{align} 
    \gamma = -\frac{1}{\pi} \int \limits_{E_{p} < V_{0} - m} dp \frac{p}{E_{p}} \ln R^{2} (p) = -\frac{1}{\pi} \int_{m}^{V_{0} - m} dE_{p} \ln R^{2} (p), \quad \mathcal{N}^{2} = \left| {}_{\mathrm{in}} \! \braket{0|0}_{\mathrm{out}} \right|^{2} = e^{-\gamma T_{\mathrm{total}}}. 
  \end{align}

  The above expressions \eqref{eq:invac_writtenby_outvac(Sec5)}--\eqref{eq:Bout(Sec5)} explicitly show that the ``in'' vacuum has nonvanishing overlaps with states of out-going pairs of particle and anti-particle, which implies that it is not stable in the Klein region. In order to evaluate explicitly such probabilities of finding out-going pairs of particle and anti-particle, it is instructive and helpful first to calculate the following generating functional 
  \begin{align} 
    G[a] = \bra{0} \exp \left[ -\sum_{s} \int \limits_{E_{p} < V_{0} - m} dp \, a(p, s) d(q, \tilde{s}) b(p, s) \right] \exp \left[ -\sum_{s} \int \limits_{E_{p} < V_{0} - m} dp \sqrt{\frac{E_{q}}{E_{p}}} \frac{T(p)}{R(p)} b^{\dagger} (p, s) d^{\dagger} (q, \tilde{s}) \right] \ket{0}, 
  \end{align} 
  where all operators and states are understood as ``out'' ones though the index ``out'' shall be suppressed for simplicity. Following the similar procedure as above, it is not difficult to show that $G[a]$ satisfies the functional differential equation 
  \begin{align} 
    \frac{\delta}{\delta a(p, s)} G[a] = \frac{\sqrt{\frac{E_{q}}{E_{p}}} \frac{T(p)}{R(p)}}{1 + \frac{q}{p} \sqrt{\frac{E_{p}}{E_{q}}} \frac{T(p)}{R(p)} a(p, s)} \braket{0 | d(q, \tilde{s}) d^{\dagger} (q, \tilde{s}) | 0} G[a] = \frac{\frac{q}{p} \sqrt{\frac{E_{p}}{E_{q}}} \frac{T(p)}{R(p)}}{1 + \frac{q}{p} \sqrt{\frac{E_{p}}{E_{q}}} \frac{T(p)}{R(p)} a(p, s)} \braket{0 | b(p, s) b^{\dagger} (p, s) | 0} G[a]. 
  \end{align} 
  The solution with the condition $G[0] = 1$ reads as 
  \begin{align} 
    G[a] = \exp \left[ \sum_{s} \int \limits_{E_{p} < V_{0} - m} dp \ln \left( 1 + \frac{q}{p} \sqrt{\frac{E_{p}}{E_{q}}} \frac{T(p)}{R(p)} a(p, s) \right) \braket{0 | b(p, s) b^{\dagger} (p, s) | 0} \right]. 
    \label{eq:explicitexpression_G(Sec5)}%
  \end{align} 
  Successive functional derivatives of $G$ evaluated at $a = 0$ generate amplitudes of finding out-going particle--anti-particle pairs in the ``in'' vacuum. For example, the first derivative yields the amplitude finding a single pair 
  \begin{align} 
    {}_{\mathrm{out}} \! \braket{0 | d_{\mathrm{out}} (q, \tilde{s}) b_{\mathrm{out}} (p, s) | 0}_{\mathrm{in}} = \mathcal{N} \left. \frac{\delta}{\delta a(p, s)} G[a] \right|_{a = 0} = \mathcal{N} \frac{q}{p} \sqrt{\frac{E_{p}}{E_{q}}} \frac{T(p)}{R(p)} {}_{\mathrm{out}} \! \braket{0 | b_{\mathrm{out}} (p, s) b_{\mathrm{out}}^{\dagger} (p, s) | 0}_{\mathrm{out}}. 
  \end{align} 
  The second derivative, on the other hand, yields the amplitude of the form 
  \begin{equation}\begin{split} 
    {}_{\mathrm{out}} \! \braket{0 | (db)_{1} (db)_{2} | 0}_{\mathrm{in}} &= \mathcal{N} \left. \frac{\delta^{2}}{\delta a_{1} \delta a_{2}} G[a] \right|_{a = 0} \\ 
    &= \mathcal{N} \left. \left( -\frac{\beta_{1}^{2} \delta (p_{1} - p_{2}) \delta_{s_{1}, s_{2}}}{(1 + \beta_{1} a_{1})^{2}} + \frac{\beta_{1} \beta_{2}}{(1 + \beta_{1} a_{1})(1 + \beta_{2} a_{2})} {}_{\mathrm{out}} \! \braket{0 | (bb^{\dagger}) | 0}_{\mathrm{out}} \right) {}_{\mathrm{out}} \! \braket{0 | (bb^{\dagger}) | 0}_{\mathrm{out}} \right|_{a = 0}, 
  \end{split}\end{equation} 
  where we have set $\frac{q}{p} \sqrt{\frac{E_{p}}{E_{q}}} \frac{T(p)}{R(p)} = \beta$ and introduced a shorthand notation, $(db)_{1} = d_{\mathrm{out}} (q_{1}, \tilde{s}_{1}) b_{\mathrm{out}} (p_{1}, s_{1})$ etc.. This amplitude vanishes when $p_{1} = p_{2}, \, s_{1} = s_{2}$ as it should be, for ${}_{\mathrm{out}} \! \braket{0 | (bb^{\dagger}) | 0}_{\mathrm{out}} = \delta (p - p)$. We thus may write 
  \begin{align} 
    \frac{\delta^{2}}{\delta a_{1} \delta a_{2}} G[a] = \begin{cases} \dfrac{\beta_{1} \, {}_{\mathrm{out}} \! \braket{0 | (bb^{\dagger}) | 0}_{\mathrm{out}}}{1 + \beta_{1} a_{1}} \dfrac{\beta_{2} \, {}_{\mathrm{out}} \! \braket{0 | (bb^{\dagger}) | 0}_{\mathrm{out}}}{1 + \beta_{2} a_{2}} G[a] & \text{for $1 \neq 2$} \\ 0 & \text{otherwise} \end{cases}, 
  \end{align} 
  which can be generalized to higher-order derivatives and we understand that the amplitude of finding $n$ pairs is given by 
  \begin{align} 
    {}_{\mathrm{out}} \! \braket{0 | (db)_{1} (db)_{2} \cdots (db)_{n} | 0}_{\mathrm{in}} = \mathcal{N} \prod_{k = 1}^{n} \beta_{k} \, {}_{\mathrm{out}} \! \braket{0 | (bb^{\dagger})_{k} | 0}_{\mathrm{out}} = \mathcal{N} \prod_{k = 1}^{n} \frac{q_{k}}{p_{k}} \sqrt{\frac{E_{p_{k}}}{E_{q_{k}}}} \frac{T(p_{k})}{R(p_{k})} {}_{\mathrm{out}} \! \braket{0 | b(p_{k}, s_{k}) b^{\dagger} (p_{k}, s_{k}) | 0}_{\mathrm{out}}, 
  \end{align} 
  where all momenta $p_{k}$ are different, for otherwise it vanishes.

  In order to properly normalize the multi-pair states, so that their inner products are just given by the product of delta functions, each state has to be devided by $\sqrt{\frac{q}{p} \frac{E_{p}}{E_{q}} {}_{\mathrm{out}} \! \braket{0 | b(p, s) b^{\dagger} (p, s) | 0}_{\mathrm{out}}}$ because 
  \begin{align} 
    {}_{\mathrm{out}} \! \braket{0 | d_{\mathrm{out}} (q, \tilde{s}) b_{\mathrm{out}} (p, s) b_{\mathrm{out}}^{\dagger} (p', s') d_{\mathrm{out}}^{\dagger} (q', \tilde{s}') | 0}_{\mathrm{out}} = \frac{q}{p} \frac{E_{p}}{E_{q}} {}_{\mathrm{out}} \! \braket{0 | b_{\mathrm{out}} (p, s) b_{\mathrm{out}}^{\dagger} (p, s) | 0}_{\mathrm{out}} \delta (p - p') \delta_{s, s'}. 
  \end{align} 
  The probability of finding an out-going single pair thus reads as 
  \begin{align} 
    P_{1} = \mathcal{N}^{2} \sum_{s} \int \limits_{E_{p} < V_{0} - m} dp \frac{q}{p} \frac{T^{2} (p)}{R^{2} (p)} {}_{\mathrm{out}} \! \braket{0 | b_{\mathrm{out}} (p, s) b_{\mathrm{out}}^{\dagger} (p, s) | 0}_{\mathrm{out}}. 
  \end{align} 
  Generally speaking, however, it is not easy to write down the probability of finding $n$ pairs of out-going particle and anti-particle, denoted as $P_{n}$, because a naive expectation for $P_{n}$, 
  \begin{align} 
    \mathcal{N}^{2} \frac{1}{n!} \left( \sum_{s} \int \limits_{E_{p} < V_{0} - m} dp \frac{q}{p} \frac{T^{2} (p)}{R^{2} (p)} {}_{\mathrm{out}} \! \braket{0 | b_{\mathrm{out}} (p, s) b_{\mathrm{out}}^{\dagger} (p, s) | 0}_{\mathrm{out}} \right)^{n}, \quad n = 0, 1, 2, \cdots, 
  \end{align} 
  is not precise and has to be corrected by properly subtracting all possible coincident contributions where at least two pairs share the same momentum. The explicit form of $P_{n}$ could instead be read from the power expansion of the generating functional $G[a]$ \eqref{eq:explicitexpression_G(Sec5)}.

\section{\label{Sec6} Scattering process seen as transition from single-particle ``in'' state} %

  Even though the asymptotic ``in'' state is shown to decay into pairs of particle and anti-particle and it is not clear whether to discuss stationary scattering process within this framework is meaningful, we can explore what happens to a single-particle state prepared at $t = -\infty$. Such a state would be interpreted as an initial state of the scattering problem. Consider, for definiteness, a state that corresponds to a left-incident particle moving right at $t = -\infty$, which is represented by the state $b_{\mathrm{in}}^{\dagger} (p, s) \ket{0}_{\mathrm{in}}$. This state is shown to contain an out-going particle and multi-pairs of particle and anti-particle at $t = \infty$. We just rewrite the initial state in terms of the ``out'' operators and ``out'' vacuum 
  \begin{align} 
    b_{\mathrm{in}}^{\dagger} (p, s) \ket{0}_{\mathrm{in}} = \left[ R(p) b_{\mathrm{out}}^{\dagger} (p, s) + \sqrt{\frac{E_{q}}{E_{p}}} T(p) d_{\mathrm{out}} (q, \tilde{s}) \right] \mathcal{N} e^{-B_{\mathrm{out}}^{\dagger}} \ket{0}_{\mathrm{out}} = \frac{\mathcal{N}}{R(p)} b_{\mathrm{out}}^{\dagger} (p, s) e^{-B_{\mathrm{out}}^{\dagger}} \ket{0}_{\mathrm{out}}, 
    \label{eq:instate_writtenby_outstates(Sec6)}%
  \end{align} 
  where the last equality follows if the operators in the parentheses are moved just next to the vacuum. Since the state contains only those states that contain one more particles than anti-particles, we just concentrate on the transition amplitudes to such states.

  It is possible to evaluate the following amplitude, which will turn out to be relevant to the current problem, to obtain 
  \begin{align} 
    {}_{\mathrm{out}} \! \bra{0} \exp \left[ -\sum_{s} \int \limits_{E_{p} < V_{0} - m} dp \, a(p, s) d_{\mathrm{out}} (q, \tilde{s}) b_{\mathrm{out}} (p, s) \right] b_{\mathrm{out}} (p', s') b_{\mathrm{out}}^{\dagger} (p, s) e^{-B_{\mathrm{out}}^{\dagger}} \ket{0}_{\mathrm{out}} = \frac{\delta (p - p') \delta_{s, s'}}{1 + \frac{q}{p} \sqrt{\frac{E_{p}}{E_{q}}} \frac{T(p)}{R(p)} a(p, s)} G[a], 
    \label{eq:generatingft_scatteringamplitudes(Sec6)}%
  \end{align} 
  where $G[a]$ is given in \eqref{eq:explicitexpression_G(Sec5)}. If we set $a(p, s) = \sqrt{\frac{E_{q}}{E_{p}}} \frac{T(p)}{R(p)}$, $G$ becomes ${}_{\mathrm{out}} \! \braket{0 | e^{-B_{\mathrm{out}}} e^{-B_{\mathrm{out}}^{\dagger}} | 0}_{\mathrm{out}} = \mathcal{N}^{-2}$ and the above relation just implies that 
  \begin{align} 
    {}_{\mathrm{out}} \! \braket{0 | e^{-B_{\mathrm{out}}} b_{\mathrm{out}} (p', s') b_{\mathrm{out}}^{\dagger} (p, s) e^{-B_{\mathrm{out}}^{\dagger}} | 0}_{\mathrm{out}} = \mathcal{N}^{-2} R^{2} (p) \delta (p - p') \delta_{s, s'}. 
    \label{eq:normalization_instates(Sec6)}%
  \end{align} 
  Observe also that 
  \begin{equation}\begin{split} 
    &{}_{\mathrm{out}} \! \bra{0} \exp \left[ -\sum_{s} \int \limits_{E_{p} < V_{0} - m} dp \, a(p, s) d_{\mathrm{out}} (q, \tilde{s}) b_{\mathrm{out}} (p, s) \right] d_{\mathrm{out}}^{\dagger} (q', \tilde{s}') \\ 
    &= a(p', s') \frac{q'}{p'} \frac{E_{p'}}{E_{q'}} {}_{\mathrm{out}} \! \bra{0} \exp \left[ -\sum_{s} \int \limits_{E_{p} < V_{0} - m} dp \, a(p, s) d_{\mathrm{out}} (q, \tilde{s}) b_{\mathrm{out}} (p, s) \right] b_{\mathrm{out}} (p', s'), 
  \end{split}\end{equation} 
  implying that the states with anti-particles one less than particles is related to those with particles one more than anti-particles.

  We expect that the initial state with a single particle moving right with momentum $p$ and spin $s$, $b_{\mathrm{in}}^{\dagger} (p, s) \ket{0}_{\mathrm{in}}$, corresponds to a final state with a single particle moving left and multi-pairs of particle and anti-particle because the former is written as in \eqref{eq:instate_writtenby_outstates(Sec6)}. The probability of finding one out-going particle with no pairs of particle and anti-particle is simply proportional to the matrix element squared 
  \begin{align} 
    \left| {}_{\mathrm{out}} \! \braket{0 | b_{\mathrm{out}} (p', s') b_{\mathrm{in}}^{\dagger} (p, s) | 0}_{\mathrm{in}} \right|^{2} = \frac{\mathcal{N}^{2}}{R^{2} (p)} \left( \delta (p - p') \delta_{s, s'} \right)^{2}. 
  \end{align} 
  Though the quantity still has to be properly normalized in order to be interpreted as a probability density, we understand that such a probability becomes exponentially small, owing to the fact that the ``in'' vacuum is not stable and decays out into pairs of particle and anti-particle. If we take into account of such decay products of particle--anti-particle pairs, the matrix element squared for a specific final state with a particle and multi-pairs of particle and anti-particle has to be summed over all degrees of freedom, i.e., we have to sum over spins and integrate over momenta of the pairs, after properly normalizing the final state. Specifically, we calculate 
  \begin{align} 
    \sum_{n \geq 0} \frac{1}{n!} \prod_{k = 1}^{n} \sum_{s_{k}} \int \limits_{E_{p_{k}} < V_{0} - m} \frac{dp_{k}}{\xi_{k}} \left| {}_{\mathrm{out}} \! \bra{0} (db)_{1} \cdots (db)_{n} b_{\mathrm{out}} (p', s') \frac{\mathcal{N}}{R(p)} b_{\mathrm{out}}^{\dagger} (p, s) e^{-B_{\mathrm{out}}^{\dagger}} \ket{0}_{\mathrm{out}} \right|^{2}, 
    \label{eq:reflectionprob_with_pairproduction(Sec6)}%
  \end{align} 
  where the shorthand notation $(db)_{k} = d_{\mathrm{out}} (q_{k}, \tilde{s}_{k}) b_{\mathrm{out}} (p_{k}, s_{k})$ has again been introduced and $\xi_{k}$ is the normalization factor 
  \begin{align} 
    {}_{\mathrm{out}} \! \braket{0 | (db)_{k} (b^{\dagger} d^{\dagger})_{l} | 0}_{\mathrm{out}} = \frac{q_{k}}{p_{k}} \frac{E_{p_{k}}}{E_{q_{k}}} {}_{\mathrm{out}} \! \braket{0 | b_{\mathrm{out}} (p_{k}, s_{k}) b_{\mathrm{out}}^{\dagger} (p_{k}, s_{k}) | 0}_{\mathrm{out}} \delta (p_{k} - p_{l}) \delta_{s_{k}, s_{l}} \equiv \xi_{k} \delta (p_{k} - p_{l}) \delta_{s_{k}, s_{l}}. 
  \end{align} 
  We note that the matrix element can be written as an $n$th functional derivative of \eqref{eq:generatingft_scatteringamplitudes(Sec6)} evaluated at $a = 0$ 
  \begin{align} 
    {}_{\mathrm{out}} \! \braket{0 | (db)_{1} \cdots (db)_{n} b_{\mathrm{out}} (p', s') b_{\mathrm{out}}^{\dagger} (p, s) e^{-B_{\mathrm{out}}^{\dagger}} | 0}_{\mathrm{out}} = \frac{-\delta}{\delta a_{1}} \cdots \frac{-\delta}{\delta a_{n}} \left[ \frac{\delta (p - p') \delta_{s, s'}}{1 + \frac{q}{p} \sqrt{\frac{E_{p}}{E_{q}}} \frac{T(p)}{R(p)} a(p, s)} G[a] \right]_{a = 0}, 
  \end{align} 
  where $a_{k} = a(p_{k}, s_{k})$ and therefore the above quantity \eqref{eq:reflectionprob_with_pairproduction(Sec6)}, which is now written as 
  \begin{equation}\begin{split} 
    \frac{\mathcal{N}^{2}}{R^{2} (p)} \sum_{n} \frac{1}{n!} \prod_{k = 1}^{n} \sum_{s_{k}} \int \limits_{E_{p_{k}} < V_{0} - m} dp_{k} \, {}_{\mathrm{out}} \bra{0} e^{-B_{\mathrm{out}}} b_{\mathrm{out}} (p, s) b_{\mathrm{out}}^{\dagger} (p', s') \frac{(b^{\dagger} d^{\dagger})_{n}}{\xi_{n}} \cdots \frac{(b^{\dagger} d^{\dagger})_{1}}{\xi_{1}} \ket{0}_{\mathrm{out}} \\ 
    \times \frac{-\delta}{\delta a_{1}} \cdots \frac{-\delta}{\delta a_{n}} \left[ \frac{\delta (p - p') \delta_{s, s'}}{1 + \frac{q}{p} \sqrt{\frac{E_{p}}{E_{q}}} \frac{T(p)}{R(p)} a(p, s)} G[a] \right]_{a = 0}, 
  \end{split}\end{equation} 
  is reduced to 
  \begin{align} 
    \frac{\mathcal{N}^{2}}{R^{2} (p)} {}_{\mathrm{out}} \! \bra{0} e^{-B_{\mathrm{out}}} b_{\mathrm{out}} (p, s) b_{\mathrm{out}}^{\dagger} (p', s') \left[ \frac{\delta (p - p') \delta_{s, s'}}{1 + \frac{q}{p} \sqrt{\frac{E_{p}}{E_{q}}} \frac{T(p)}{R(p)} a(p, s)} G[a] \right]_{a = -\frac{(b^{\dagger} d^{\dagger})}{\xi}} \ket{0}_{\mathrm{out}}. 
  \end{align} 
  Since $b_{\mathrm{out}}^{\dagger}$ and $d_{\mathrm{out}}^{\dagger}$ are fermion operators that satisfy $(b_{\mathrm{out}}^{\dagger} (p, s))^{2} = 0 = (d_{\mathrm{out}}^{\dagger} (q, \tilde{s}))^{2}$, the operator inserted in the denominator gives no contribution and we have 
  \begin{align} 
    G[a] \biggr|_{a = -\frac{(b^{\dagger} d^{\dagger})}{\xi}} = \exp \left[ -\sum_{s} \int \limits_{E_{p} < V_{0} - m} dp \sqrt{\frac{E_{q}}{E_{p}}} \frac{T(p)}{R(p)} b_{\mathrm{out}}^{\dagger} (p, s) d_{\mathrm{out}}^{\dagger} (q, \tilde{s}) \right], 
  \end{align} 
  which is nothing but $e^{-B_{\mathrm{out}}^{\dagger}}$. The quantity \eqref{eq:reflectionprob_with_pairproduction(Sec6)} finally turns out to be 
  \begin{align} 
    \frac{\mathcal{N}^{2}}{R^{2} (p)} {}_{\mathrm{out}} \! \braket{0 | e^{-B_{\mathrm{out}}} b_{\mathrm{out}} (p, s) b_{\mathrm{out}}^{\dagger} (p', s') e^{-B_{\mathrm{out}}^{\dagger}} | 0}_{\mathrm{out}} \delta (p - p') \delta_{s, s'} = \delta (p - p') \delta_{s, s'} \delta (p - p). 
    \label{eq:calculation_reflectionprob(Sec6)}%
  \end{align} 
  The last equality follows from the normalization condition \eqref{eq:normalization_instates(Sec6)}.

  In order to properly reproduce probabilities or ``cross section'' in scattering problem, the result \eqref{eq:calculation_reflectionprob(Sec6)} has to be devided by the incident flux and the total scattering time. The incident flux $j_{\mathrm{inc}} = \psi_{s}^{\dagger} (p) \alpha_{z} \psi_{s} (p)$ is calculated from the incident wave function 
  \begin{align} 
    \psi_{s} (p) = \frac{1}{\sqrt{2\pi}} \sqrt{\frac{m}{E_{p}}} u(p, s) 
  \end{align} 
  and we obtain 
  \begin{align} 
    j_{\mathrm{inc}} = \frac{1}{2\pi} \frac{m}{E_{p}} \frac{E_{p} + m}{2m} \frac{2p}{E_{p} + m} = \frac{1}{2\pi} \frac{p}{E_{p}}. 
  \end{align} 
  Since the total scattering time $T_{\mathrm{total}}$ is formally written as 
  \begin{align} 
    T_{\mathrm{total}} = 2\pi \delta (E_{p} - E_{p}) = 2\pi \frac{E_{p}}{p} \delta (p - p), 
  \end{align} 
  the amplitude squared has to be multiplied by 
  \begin{align} 
    \frac{1}{T_{\mathrm{total}} \, j_{\mathrm{inc}}} = \frac{1}{\delta (p - p)} 
  \end{align} 
  to yield the probability density. Notice that the same result could be attained if the initial state is normalized just by formally dividing it by ${}_{\mathrm{in}} \! \braket{0 | b_{\mathrm{in}} (p, s) b_{\mathrm{in}}^{\dagger} (p, s) | 0}_{\mathrm{in}} = \delta (p - p)$.

  We arrive at a conclusion that when a particle with momentum $p > 0$ and spin $s$ is injected, the probability of finding an out-going particle moving left asymptotically is just one, when one does not care about the number of particle--anti-particle pairs produced 
  \begin{align} 
    \text{Prob. (particle moving right with $p, s$ $\to$ particle moving left with arbitrary numbers of pairs)} = 1. 
  \end{align} 
  Since the out-going particle moves left, this is considered to be the reflection probability, i.e., total reflection and states that this is all what happens.

\section{\label{Sec7} Summary and discussions} %

  The Klein tunneling process in relativistic quantum mechanics is examined on the basis of quantum field theory, where the field operator $\Psi$ which is assumed to have all information on the physical process plays crucial and central roles. It is constructed on the basis of the solutions of the Dirac equation that satisfy boundary conditions for stationary scattering problem. It is stressed that the condition that they constitute a complete orthonormal set is crucial in order for operators introduced as the expansion coefficients to satisfy the standard anti-commutation relations. In constrast to most cases where the completeness of the basis functions has just been assumed on physical ground, the stationary solutions of the Dirac equation on which the field quantization of this paper is based have been shown explicitly to satisfy the completeness condition \cite{Ochiai2018}. This fact guarantees the standard anti-commutation relations among creation/annihilation operators introduced in the field operator. Other choices of basis functions, of course, would be possible, but only when they are shown (not simply assumed) to be complete and orthonormal.

  In this respect, it should be remarked that the orthogonality of the stationary solutions belonging to the same energy eigenvalue is closely connected to the choice of proper boundary conditions. As stated above and can be shown explicitly in simple cases, the usual boundary condition for the scattering process, that is, the existence of incident and reflected plane waves on one far side of the interaction region and the transmitted plane wave on the other far side, entails, in general, to the orthogonality of left-incident and right-incident solutions. The result can be derived in connection with a position-independent current formed from these solutions. It should be observed that the other solutions that satisfy, say, spatially asymptotic plane wave conditions, that is, solutions that have either out-going or in-coming plane waves on the same far side of the interaction region, are not orthogonal to each other. See Appendix \ref{AppendB} for details.

  Next, the asymptotic operators are introduced as the appropriate $t = \pm \infty$ limits of the field operator $\Psi$. At $t = -\infty$, only the right-moving wave exsisting at the left of the step ($z < 0$) and left-moving wave in the potential region ($z > 0$) survive in $\Psi$, which are supposed to represent incident waves. On the other hand, only the left-moving and right-moving waves are extracted for $z < 0$ and $z > 0$, respectively, at $t = \infty$, representing out-gonig scattered ones. The asymptotic operators are just defined as the coefficients of these waves in $\Psi$. The asymptotic vacuum states are then defined to be those annihilated by the asymptotic annihilation operators. Two sets of asymptotic operators are related by the Bogoliubov-like transformation in the Klein energy range, on which we have exclusively forcused in this paper, resulting in an unstable ``in'' vacuum.

  In this way, we arrive at the following view on the Klein tunneling on the basis of quantum field theory. When the step potential is higher than $2m$, not only the ``in'' vacuum is unstable, but also a particle injected into such a potential is totally reflected and is accompanied with pair-produced particles moving left and anti-particles moving right. No transmitted particles moving right exist. This view would have been somewhat expected when the field operator $\Psi$ is expanded in terms of $b_{\mathrm{out}}$ and $d_{\mathrm{out}}^{\dagger}$, however, the explicit calculations presented in the previous sections show consistently that it is indeed the case.

  Even though in this paper we have exclusively considered operators relevant to the Klein region, the framework developed here surely admits presence of other particle and anti-particle modes, for which both asymptotic vacua are the same because no mixing between creation and annihilation operators is necessary to connect different asymptotic operators. Nothing peculiar would happen for such modes.

\appendix 
\section{\label{AppendA} Dirac spinors} %

  We fix the notations and normalizations of four-component spinors. The stationary plane-wave solutions for the free Dirac Hamiltonian 
  \begin{align} 
    H_{0} = -i\alpha_{z} \partial_{z} + \beta m, \quad \alpha_{z} = \begin{pmatrix} 0 & \sigma_{z} \\ \sigma_{z} & 0 \end{pmatrix}, \quad \beta = \begin{pmatrix} \1 & 0 \\ 0 & -\1 \end{pmatrix} 
  \end{align} 
  are 
  \begin{align} 
    u(p, s) e^{-ip \cdot z} = \sqrt{\frac{E_{p} + m}{2m}} \begin{pmatrix} \1 \\ \frac{\sigma_{z} p}{E_{p} + m} \end{pmatrix} \bm{\xi} (s) e^{ipz - iE_{p} t} 
  \end{align} 
  for a positive frequency $E_{p} = \sqrt{p^{2} + m^{2}}$ and 
  \begin{align} 
    v(p, s) e^{ip \cdot z} = \sqrt{\frac{E_{p} + m}{2m}} \begin{pmatrix} \frac{\sigma_{z} p}{E_{p} + m} \\ \1 \end{pmatrix} \bm{\xi} (s) e^{-ipz + iE_{p} t} 
  \end{align} 
  for a negative frequency $-E_{p}$, where $\sigma_{z}$ is the third Pauli matrix, $\1$ the $2\times 2$ unit matrix and $\bm{\xi} (s)$ a two-component spinor. They are normalized and form a complete orthonormal set 
  \begin{gather} 
    \bar{u} (p, s) u(p, s') = \delta_{s, s'}, \quad \bar{v} (p, s) v(p, s') = -\delta_{s, s'}, \label{eq:normalization_spinor(AppendA)} \\ 
    u^{\dagger} (p, s) u(p, s') = v^{\dagger} (p, s) v(p, s') = \frac{E_{p}}{m} \delta_{s, s'}, \\ 
    \sum_{s} \left[ u(p, s) \bar{u} (p, s) - v(p, s) \bar{v} (p, s) \right] = \1_{4\times 4}. \label{eq:completeness_spinor(AppendA)} 
  \end{gather} 
  Now, if a constant potential is added to the above Hamiltonian 
  \begin{align} 
    H_{0}' = H_{0} + V_{0} = -i\alpha_{z} \partial_{z} + \beta m + V_{0} \quad (-\infty < z < \infty), 
  \end{align} 
  the corresponding plane-wave solutions are ($E_{q} = \sqrt{q^{2} + m^{2}}$) 
  \begin{align} 
    u(q, s) e^{-iq \cdot z} = \sqrt{\frac{E_{q} + m}{2m}} \begin{pmatrix} \1 \\ \frac{\sigma_{z} q}{E_{q} + m} \end{pmatrix} \bm{\xi} (s) e^{iqz - i(E_{q} + V_{0}) t} 
  \end{align} 
  for a ``positive'' frequency $E_{q} + V_{0} \geq V_{0}$ and 
  \begin{align} 
    v(q, s) e^{iq \cdot z} = \sqrt{\frac{E_{q} + m}{2m}} \begin{pmatrix} \frac{\sigma_{z} q}{E_{q} + m} \\ \1 \end{pmatrix} \bm{\xi} (s) e^{-iqz - i(-E_{q} + V_{0}) t} 
  \end{align} 
  for a ``negative'' frequency $-E_{q} + V_{0} \leq V_{0}$. They are again normalized and form a complete orthonormal set in spinor space, just like in \eqref{eq:normalization_spinor(AppendA)}--\eqref{eq:completeness_spinor(AppendA)}, with $E_{p}$ replaced with $E_{q}$. Notice that the frequency $-E_{q} + V_{0}$ can be positive for small momenta $|q| < \sqrt{V_{0}^{2} - m^{2}}$, even though it is to be associated with the spinor $v$. Note also that the above explicit forms of spinors are not unique since $\sigma_{z}$ is commutable with the above $H_{0}$ and $H_{0}'$ and therefore, e.g., $\sigma_{z} u(p, s)$ can be used instead of $u(p, s)$.

  Incidentally, the current conservation $\partial_{0} j^{0} + \partial_{z} j^{z} = 0$, resulting from the invariance under the global phase transformation, implies a $z$-independent current $j^{z} = \psi^{\dagger} \alpha_{z} \psi$ for stationary states and the above positive- and negative-frequency solutions give both positive currents 
  \begin{align} 
    u^{\dagger} (q, s) \sigma_{z} u(q, s) = \frac{q}{m} = v^{\dagger} (q, s) \sigma_{z} v(q, s) 
  \end{align} 
  for positive $q > 0$, irrespectively of the value of $V_{0}$. In this respect, both $u(q, s)$ and $v(q, s)$ are considered to express positive flows of current with positive and negative frequencies.

\section{\label{AppendB} Orthogonality and boundary conditions of stationary solutions} %

  We show here that the orthogonality of stationary solutions of the Dirac equation that belong to the same energy eigenvelue is closely connected to the boundary conditions they satisfy and the existence of spatially constant quantities. First, under the standard boundary conditions for scattering problems, a left-incident solution $\psi$ is characterized by its asymptotic behavior 
  \begin{align} 
    z \to -\infty: \quad \psi \sim e^{\pm ipz} \chi_{p} + R_{\gets} \chi_{-p} e^{\mp ipz}, \qquad z \to \infty: \quad \psi \sim T_{\to} \chi_{p'} e^{\pm ip'z}, 
    \label{eq:bc_psi(AppendB)}%
  \end{align} 
  and a right-incident solution $\phi$ by 
  \begin{align} 
    z \to -\infty: \quad \phi \sim T_{\gets} e^{\mp ipz} \chi_{-p}, \qquad z \to \infty: \quad \phi \sim e^{\mp ip'z} \chi_{-p'} + R_{\to} \chi_{p'} e^{\pm ip'z}, 
    \label{eq:bc_phi(AppendB)}%
  \end{align} 
  where $\chi$ represents either spinor $u$ (positive frequency solution) or $v$ (negative frequency solution), upper and lower signs in the exponent are to be associated with $u$ and $v$, respectively (spin degrees of freedom is neglected) and $p > 0$ and $p' > 0$ are magnitudes of asymptotic momenta at $z = -\infty$ and $z = \infty$ where the potential takes constant values. Since both $\psi$ and $\phi$ satisfy the same Dirac equation with the same energy, the quantity $\psi^{\dagger} \alpha_{z} \phi$ is shown to be $z$-independent $\partial_{z} (\psi^{\dagger} \alpha_{z} \phi) = 0$, in particular, its values at $z = \pm \infty$ are the same. Inserting their spatially asymptotic forms \eqref{eq:bc_psi(AppendB)} and \eqref{eq:bc_phi(AppendB)}, we have 
  \begin{align} 
    \psi^{\dagger} \alpha_{z} \phi \sim \left( e^{\mp ipz} \chi_{p}^{\dagger} + R_{\gets}^{*} e^{\pm ipz} \chi_{-p}^{\dagger} \right) \alpha_{z} T_{\gets} e^{\mp ipz} \chi_{-p} = -\frac{p}{m} R_{\gets}^{*} T_{\gets} 
  \end{align} 
  at $z = -\infty$ and 
  \begin{align} 
    \psi^{\dagger} \alpha_{z} \phi \sim T_{\to}^{*} e^{\mp ip'z} \chi_{p'}^{\dagger} \alpha_{z} \left( e^{\mp ip'z} \chi_{-p'} + R_{\to} e^{\pm ip'z} \chi_{p'} \right) = \frac{p'}{m} T_{\to}^{*} R_{\to} 
  \end{align} 
  at $z = \infty$, resulting in the equality (reciprocity) 
  \begin{align} 
    -\frac{p}{m} R_{\gets}^{*} T_{\gets} = \frac{p'}{m} T_{\to}^{*} R_{\to}. 
    \label{eq:reciprocity(AppendB)}%
  \end{align} 
  Here we have used the relation $\chi_{p}^{\dagger} \alpha_{z} \chi_{p} = \frac{p}{m}$ and $\chi_{p}^{\dagger} \alpha_{z} \chi_{-p} = 0$ that hold for any choice of $\chi$.

  Now consider an inner product between $\psi_{1}$ with left-incident momentum $p_{1}$ and $\phi_{2}$ with right-incident momentum $p_{2}'$. If $p_{1} \neq p_{2}$ where $p_{2}$ is the asymptotic momentum corresponding to $p_{2}'$, two solutions are belonging to different energy eigenvalues and are thus orthogonal to each other. We thus anticipate that if two momenta become identical, only singular contributions proportional to delta functions would appear in the inner product (for any finite contributions are absent if the momenta are different). Such contributions only appear from the spatially asymptotic regions $z \sim \infty$ and $z \sim -\infty$ and can be estimated as 
  \begin{equation}\begin{split} 
    \int_{-\infty}^{\infty} dz \psi_{1}^{\dagger} \phi_{2} &= \text{divergent parts of} \left[ \int_{-\infty}^{z_{-}} dz \left( e^{\mp ip_{1} z} \chi_{p_{1}}^{\dagger} + R_{\gets}^{*} e^{\pm ip_{1} z} \chi_{-p_{1}}^{\dagger} \right) T_{\gets} e^{\mp ip_{2} z} \chi_{-p_{2}} \right. \\ 
    &\phantom{= \text{divergent parts of }} \left. + \int_{z_{+}}^{\infty} dz \, T_{\to}^{*} e^{\mp ip_{1}' z} \chi_{p_{1}'}^{\dagger} \left( e^{\mp ip_{2}' z} \chi_{-p_{2}'} + R_{\to} e^{\pm ip_{2}' z} \chi_{p_{2}'} \right) \right] \\ 
    &= \pi R_{\gets}^{*} T_{\gets} \frac{E_{p_{1}}}{m} \delta (p_{1} - p_{2}) + \pi T_{\to}^{*} R_{\to} \frac{E_{p_{1}'}}{m} \delta (p_{1}' - p_{2}') = 0, 
  \end{split}\end{equation} 
  where $z_{+} > 0$ and $z_{-} < 0$ are arbitrary large and small but finite numbers. In this estimation, rapidly oscillating terms give no contributions owing to the Riemann--Lebesgue lemma and use has been made of $p_{1}' E_{p_{1}} \delta (p_{1} - p_{2}) = p_{1} E_{p_{1}'} \delta (p_{1}' - p_{2}')$ which follows from the equality $E_{p_{1}} - E_{p_{2}} = E_{p_{1}'} - E_{p_{2}'}$ and of \eqref{eq:reciprocity(AppendB)}. The orthogonality between left-incident and right-incident degenerate solutions of the stationary Dirac equation is thus proved.

  Consider next another degenerate solution of the same Dirac equation but with different boundary condition $\tilde{\phi}$ that has only left-moving component at the far right end $z \sim \infty$. It should behave asymptotically like 
  \begin{align} 
    z \to -\infty: \quad \tilde{\phi} \sim Ae^{\pm ipz} \chi_{p} + Be^{\mp ipz} \chi_{-p}, \qquad z \to \infty: \quad \tilde{\phi} \sim e^{\mp ip'z} \chi_{-p'}. 
  \end{align} 
  Following the same line of thought, a spatially conserved quantity $\psi^{\dagger} \alpha_{z} \tilde{\phi}$ derives the equality 
  \begin{align} 
    \frac{p}{m} (A - R_{\gets}^{*} B) = 0. 
  \end{align} 
  The inner product between $\psi_{1}$ and $\tilde{\phi}_{2}$ is similarly evaluated as 
  \begin{align} 
    \int_{-\infty}^{\infty} dz \psi_{1}^{\dagger} \tilde{\phi}_{2} \propto \pi (A + R_{\gets}^{*} B) \frac{E_{p_{1}}}{m} \delta (p_{1} - p_{2}) = 2\pi A \frac{E_{p_{1}}}{m} \delta (p_{1} - p_{2}), 
  \end{align} 
  which does not vanish because none of $A$ and $B$ are allowed to vanish. A similar argument is also applied to cases with two solutions that have a single component in the other asymptotic region $z \to -\infty$ and we can show that such solutions are not orthogonal to each other. We thus conclude that the degenerate solutions of the stationary Dirac equation, one with only right-moving component and the other with only left-moving component in the same spatial asymptotic region, are never orthogonal to each other.

\bibliographystyle{prsty}

\begin{thebibliography}{99}
\bibitem{comment:vac-to-vac} An essentially the same formula for vacuum decay is seen in \cite{Gavrilov2016}, even though their formalism is different from that in this paper.
\end{thebibliography}

\end{document}